\documentclass[a4paper,fleqn,usenatbib]{mnras}

\usepackage[T1]{fontenc}
\usepackage{ae,aecompl}
\usepackage{graphicx}	% Including figure files
\usepackage{amsmath}	% Advanced maths commands
\usepackage{amssymb}	% Extra maths symbols
\usepackage{pdfpages}   % include pdf 
\usepackage{acro}       % DeclareAcronym
\usepackage{cleveref}
\usepackage{setspace}

\usepackage{newtxtext,newtxmath}

\DeclareAcronym{gr}{
  short = GR ,
  long = General Relativity ,
  short-plural = ,
}

\DeclareAcronym{smbh}{
  short = SMBH ,
  long = Supermassive Black Hole ,
  short-plural = s,
}
\DeclareAcronym{gw}{
  short = GW ,
  long = Gravitational Wave ,
  short-plural = ,
}

\DeclareAcronym{lisa}{
  short = {\it LISA} ,
  long = {\it Laser Interferometric  Space  Antenna},
  short-plural = ,
}

\DeclareAcronym{pn}{
  short = PN ,
  long = post-Newtonian ,
  short-plural = ,
}
\DeclareAcronym{bh}{
  short = BH ,
  long = Black Hole ,
  short-plural = s,
  long-plural = s,
}

\DeclareAcronym{ligo}{
  short = LIGO ,
  long = Laser Interferometer Gravitational-Wave Observatory ,
  short-plural = ,
}
\DeclareAcronym{msp}{
  short = MSP ,
  long = millisecond pulsar ,
  short-plural = s,
  long-plural = s,
}
\DeclareAcronym{emrb}{
  short = EMRB ,
  long = extreme-mass-ratio binary ,
  short-plural = s,
  long-plural-form = extreme-mass-ratio binaries,
}
\DeclareAcronym{emri}{
  short = EMRI ,
  long = extreme-mass-ratio-inspiral ,
  short-plural = s,
}
\DeclareAcronym{mpd}{
  short = MPD ,
  long = Mathisson-Papapetrou-Dixon ,
  short-plural = ,  
}
\DeclareAcronym{eom}{
  short = EOMs ,
  long = equations of motion ,
}
\DeclareAcronym{gc}{
  short = GC ,
  long = globular cluster ,
  short-plural = s,
  long-plural = s,
}
\DeclareAcronym{ssc}{
  short = SSC ,
  long = spin supplementary condition ,
  short-plural = s,
  long-plural = s,
}
\DeclareAcronym{gm}{short=GM,long=gravito-magnetic}

\title[Gravito-magnetic clock effects]
{Generic gravito-magnetic clock effects}

\author[K. J. Li et al.]{
Kaye Jiale Li$^{1}$\thanks{E-mail: j-li.19@ucl.ac.uk (KJL), kinwah.wu@ucl.ac.uk (KW), z.younsi@ucl.ac.uk (ZY), joana.teixeira.19@ucl.ac.uk (JT), 
dinesh.singh@uregina.ca (DS)},
Kinwah Wu$^{1}$,
Ziri Younsi$^{1}$, 
Joana Teixeira$^{1}$, 
and Dinesh Singh$^{2}$ \\
$^{1}$Mullard Space Science Laboratory, University College London, Holmbury St Mary, Surrey, RH5 6NT, United Kingdom \\
$^{2}$Department of Physics, University of Regina, Regina, SK S4S 0A2, Canada
}

\date{Accepted 2024 March 25. Received 2024 March 13; in original form 2023 August 1}

\pubyear{2020}

\begin{document}
\label{firstpage}
\pagerange{\pageref{firstpage}--\pageref{lastpage}}
\maketitle

% Abstract of the paper
\begin{abstract}
General relativity predicts that 
 two counter-orbiting clocks 
 around a spinning mass 
 differ in the time required to complete the same orbit. 
The difference in these two values for 
 the orbital period  
 is generally referred to 
 as the \acf{gm} clock effect. 
It has been proposed 
  to measure 
  the GM clock effect using atomic clocks carried by satellites 
  in prograde and retrograde orbits 
  around the Earth. 
The precision and stability required for satellites 
  to accurately perform this measurement 
  remains a challenge for current instrumentation. 
% new system -> new formula
%Measurement of GM clock effect requires an 
%  clock of high precison and stability. 
%This naturally drives our attention to compact binaries containing a millisecond pulsar. 
One of the most accurate clocks in the Universe is a                    % DS   
  millisecond pulsar,                                                     % DS
  which emits                                                             % DS
  periodic radio pulses with high stability. 
Timing of the pulsed signals 
  from millisecond pulsars 
  has proven to be very successful in testing predictions 
  of general relativity and 
the \ac{gm} clock effect  
  is potentially measurable in binary systems. 
% what is new 
In this work 
  we derive the generic \ac{gm} clock effect by 
  considering a slowly-spinning binary system on an elliptical orbit, 
  with both arbitrary mass ratio and arbitrary spin orientations. 
The spin-orbit interaction 
  introduces a perturbation 
  to the orbit, causing the orbital plane 
  to precess and nutate. 
% short conclusion 
We identify several different contributions 
  to the clock effects: 
  the choice of spin supplementary condition
  and the observer-dependent definition of a full revolution and ``nearly-identical'' orbits. 
We discuss the impact of these subtle definitions 
  on the formula for GM clock effects
  and show that most of the existing formulae in the literature can be recovered under appropriate assumptions. 

\end{abstract}

% Select between one and six entries from the list of approved keywords.
% Don't make up new ones.

%%%%%%%%%%%%%%%%%%%%%%%%%%%%%%%%%%%%%%%%%%%%%%%%%%

%%%%%%%%%%%%%%%%% BODY OF PAPER %%%%%%%%%%%%%%%%%% 

%%%%%%%%%%%%%%%%%%%%%%%%%%%%%%%%%%%%%% 
%%%%%%%%%%%%%%%%%%%%%%%%%%%%%%%%%%%%%% 
% Sec 1 
\section{Introduction}
\label{sec:intro} 

% Introduce GEM 
In Einstein's theory of general relativity 
 a rotating mass distorts the spacetime around it, 
 leading to the precession of orbits of an object 
 around this rotating mass. 
This is known as Lense-Thirring precession \citep{Lense1918}.
It is a manifestation of frame dragging.  
An analogy to frame dragging can be found 
  by comparing the  phenomenon to electromagnetism,                          
  in which the gravitational mass is treated as equivalent to the  
  electric charge 
  and the gravitational field generated 
  by a point mass is referred to as 
  the gravito-electric field. 
The rotation of a point mass induces 
  an additional gravitational field                      
  analogous to the magnetic field produced by a spinning charge in electromagnetism, 
  and this additional gravitational field 
  corresponds to the \acf{gm} field.  
The gravito-electric field 
  gives rise to a clock effect 
  via gravitational time dilation (i.e. gravitational redshift), 
  which has been widely applied in synchronising clocks in GPS satellites. 
Such time dilation was confirmed  
   by \cite{Overstreet2022}\footnote{We note that although the term ``gravitational Aharonov–Bohm effect'' is used in their publication, it does not refer to the    % DS
   \ac{gm} potential but to the                                           
   gravito-electric potential.} experimentally,
  using atomic interference. 

The \ac{gm} field also leads                                 
  to desynchronisation of the clocks for satellites moving in prograde    
  and retrograde motion with respect to the Earth's rotation axis. 
 \cite{Cohen1988} obtained a ``synchronisation gap'' of $1.92\times 10^{-17}$ sec 
  for geosynchronous and anti-geosynchronous orbits. 
\citet{Cohen1993} suggested another possible measurement of the \ac{gm} clock effect, which arise from the  difference in time periods between a full revolution 
   of a prograde ($\mathcal{T}_+$) orbit          
  and that of 
  an otherwise equivalent retrograde ($\mathcal{T}_{-}$) orbit. 
This leads to the well known formula 
\begin{align} \label{eq:wellknown}
\mathcal{T}_+ - \mathcal{T}_- = 4 \pi \frac{J}{M [c]^2} \ , 
\end{align}
\citep[see also e.g.][]{Mashhoon1999,Tartaglia2000,Iorio2002,Faruque2004},
where $c$ is the speed of light, 
  and $J$ and $M$ are respectively 
  the spin angular momentum and mass of the Earth.
This formula is applicable to circular orbits 
  around the Earth, 
  irrespective of whether they are geosynchronous or not.
There are many interesting features of this time difference. 
The topological nature of this effect 
  is evident from its independence on the orbit's radius, 
  similar to the Aharonov-Bohm effect in quantum mechanics 
    \citep[][]{Aharonov1959}. 
This effect is expected to be about $10^{-7}$ seconds for the Earth
  \citep{Cohen1993}, 
well within the timing precision achieved by current atomic clocks 
  \citep[see][]{Mann2018}. 
It has been proposed that two satellites orbiting the Earth in prograde and retrograde motion will be able to reveal this effect.
However, the precise tracking of satellite positions,
  the dynamic part of Earth's gravity, 
  and other factors such as radiation pressure pose significant challenges
  \citep[see][]{Gronwald1997, Lichtenegger2000}. 

% why is it necessary to calculate eccentric orbit; prospects of observing this effects in pulsar systems
A promising yet often overlooked approach to detect this effect 
  is through binary pulsar systems.   
Millisecond pulsars are considered 
  as the most stable natural clocks in our universe. 
Their spin-down rates, ${\rm d}P/{\rm d}t$,  
  are in general smaller than 
  $10^{-16}\;\! {\rm s~s}^{-1}$, 
  which corresponds to a drift 
  of a few seconds over the Hubble time  
  \citep[see e.g.][]{Lorimer2008,Manchester2017}. 
The rotational stability of millisecond pulsars
  makes them a useful apparatus 
  to measure the relativistic effects 
  that are intrinsically small. 
For instance, 
  the accurate measurement of the rate of decrease 
  in the orbital period of the Hulse-Taylor binary 
  provided the first indirect proof 
  of emission of gravitational radiation 
  \citep[see][]{Taylor1989,Weisberg2005}. 
The various gravitational time dilation effects 
  have been widely used in analysing the pulsar timing data \citep[see e.g.][]{Edwards2006,Lorimer2012}.
Pulsars can also be used to test the equivalence principle \citep[see][for a review]{Stairs2003}. 
For example, the triple stellar system PSR~J0337+1715 has provided a constraint      % DS
on the strong-field violation of strong equivalence principle \citep{Shao2016,Archibald2018,Voisin2020}. 
% the 2022 paper is about a planet in this system
Similarly, \citet{Remmen2013} proposed testing general relativity with
  a hierarchical three-body system 
  consisting of a double-pulsar system revolving around a massive black hole. 
It was also demonstrated that  
  extreme-mass-ratio binaries with a fast-spinning pulsar 
  revolving around a massive black hole 
  can be used as a probe of 
  the relativistic spin-orbit coupling   
  \citep[see][]{Singh2014,Kimpson2019a,Li2019,Kimpson2020b,Li2022,Wu2022}.

% how to measure this effect of GEM clock effects
In order to measure the \ac{gm} clock effects in a pulsar binary system, 
  it is necessary to derive the formula of the clock effect for a generic orbit.     
Previous studies \citep{Cohen1993,Mashhoon1999,Tartaglia2000,Bini2001,Bini2002,Iorio2002,Faruque2004} mostly focused on circular orbits with spin aligned or anti-aligned with the orbital angular momentum, 
  such that the orbits                                                 % DS 
  remain on the equatorial plane. 
%Among them, there are mainly few different approaches. 
\citet{Teyssandier1977} and \citet{Teyssandier1978} 
  extended the study to orbits around a                                % DS
  rotating non-spherical mass. 
\citet{Faruque2003} derived a different \ac{gm} 
  clock effect formula 
  by comparing the prograde innermost stable circular orbit and the    % DS
  retrograde innermost stable circular orbit. 
The effect of the cosmological constant on the \ac{gm} 
  clock effect is studied by \citet{Kerr2003}. 
\citet{Mashhoon1999} and \citet{Iorio2002} 
  generalised the \ac{gm} clock effect to arbitrary inclination angle for 
  circular orbits. 
As well,                                                              % DS
\citet{Mashhoon2001} extended their work to eccentric 
  and inclined orbits in an Earth-satellite system, 
  under the exact \ac{gm} analogue. 
The effect                                                            % DS
of eccentricity is also taken into account by 
  \citet{Hackmann2014} by studying the motion of test particles 
  in Kerr spacetime on eccentric orbits.                              % DS 
  They obtained results that differ from those of other literature on the topic.  % DS
The effect of spin for the test mass                                  % DS
on circular (or semi-circular) orbits 
  is included by \citet{Bini2004a,Bini2004b,Faruque2006,Mashhoon2006} 
  but different coefficients were reported. 

In the work we derive a formula 
  for the general \ac{gm} clock effect 
  involving two spinning masses on an orbit with 
  arbitrary inclination and eccentricity. 
The formula is applicable over a large parameter range, 
  and we show that it can reproduce the results 
  obtained in most other previous studies 
  with appropriate assumptions. 
We also clarify the origin of apparent discrepancy 
  in the formulae reported in different papers. 
We organise the paper as follow. 
Sec.~\ref{sec:spinorbit} gives a brief introduction to 
  spin-orbit coupling with different \acp{ssc}. 
Sec.~\ref{sec:perturbation} presents 
  the \ac{eom} of 
  a binary system subjected to the spin-orbit coupling force
  and our derivation of 
  the generic formula for \ac{gm} clock effect. 
We show in Sec.~\ref{sec:literature} 
  our results are generalisation of many previous works 
  and how we may use our formula to 
  resolve the issues of the discrepancies of previous work.  
We also provide 
  an in-depth exposition 
  of the concept of \ac{gm} clock effects,  
  highlighting common misconceptions 
  and misinterpretations (of results in observations) 
  often found in literature.  
Further, we comment on implications on the 
  observation of the \ac{gm} clock effect via the pulsar binary,
  and via artificial satellite systems.  
Unless otherwise stated,  we adopt   
  the $[-,+,+,+]$ metric signature convention 
  and the natural unit system with $c = G =1$ 
  in this paper, 
  where $c$ is the speed of light and 
  $G$ is the gravitational constant. 

\section{Spin-orbit coupling} 
\label{sec:spinorbit}

\subsection{Spin-orbit coupling and spin supplementary condition}
 
Consider a binary system with masses $m_1$ and $m_2$, 
  respectively located 
  at ${\boldsymbol r}_1$ and ${\boldsymbol r}_2$,
  moving with ${\boldsymbol v}_1$ and ${\boldsymbol v}_2$ with respect to their centre-of-mass. 
We define a mass ratio $q \equiv m_2/m_1$.   
The total mass of the binary is therefore  
  $M = m_1 + m_2 = m_1(1+q)$, 
  the relative position 
  ${\boldsymbol r} = {\boldsymbol r}_1 - {\boldsymbol r}_2$ and 
  relative velocity ${\boldsymbol v} = {\boldsymbol v}_1 - {\boldsymbol v}_2$. 
It follows that the total spin 
${\boldsymbol S} = {\boldsymbol S}_1 + {\boldsymbol S}_2$,
  and the mass-weighted total spin is  
  ${\boldsymbol \sigma} = q\;\! {\boldsymbol S}_1 
  + q^{-1} {\boldsymbol S}_2 $.  
As the \ac{gm} clock effect can only be measured 
  in a system with at least one accurate clock,
  we take $m_1$ to be a pulsar (neutron star) 
  and $m_2$ be its companion star. 
There are no restrictions on the value of $q$  
  in the derivation presented in this work, 
  i.e. the companion star 
    can be less or more massive, 
    than the pulsar,  
    and hence it can a main sequence star, 
    a white dwarf, a neutron star or a black hole. 

In additional to the Newtonian gravitational 
  force $-M {\boldsymbol r}/r^3$, the  
  spin-orbit coupling force produces an additional 
  acceleration on the binary, and it is given by  
\begin{equation}
\begin{aligned} 
\label{eq:spinorbitforce}
{\boldsymbol a}_{\rm SO} = & 
    \frac{1}{r^3} \bigg\{ \frac{3}{r^2}\;\!  
     {\boldsymbol r} \;\! 
    \big[\;\! \left( {\boldsymbol r}\times 
    {\boldsymbol v}\right) \cdot \left((k+1)\;\! 
    {\boldsymbol \sigma }+2 {\boldsymbol S}\right) 
    \big]   \\    
& +\frac{3 \dot{r}}{r} \big[\;\! {\boldsymbol r}\times \left((2-k) {\boldsymbol \sigma }+2 {\boldsymbol S}\right) \big] 
-{\boldsymbol v}\times \left(3 {\boldsymbol \sigma }+4 {\boldsymbol S}\right) \bigg\}   
\end{aligned} 
\end{equation} 
\citep{Kidder1995},  
 with $k$ a parameter depending on the chosen \acf{ssc}. 
Here and hereafter, 
  ``dot'' above a variable 
  represents the differentiation with respect to time. 
It is worth noting that 
  the \ac{ssc} only affects the coefficient of ${\boldsymbol \sigma}$.
Before delving into the discussion 
  of \ac{ssc}, it is important to highlight
  that there exists a gravito-electric effect 
  (i.e. the $1$pN (post-Newtonian) effect), which is typically more significant than the spin-orbit coupling effect (a $1.5$pN effect).
However, the gravito-electric effect does not 
  contribute to $\mathcal{T}_+ - \mathcal{T}_-$
  \citep[see Eqn.~(21) of][]{Wu2022}.
It is also crucial to emphasise that the GM clock effect
  is inherently topological,
  as evidenced by the absence of orbital size 
  in the formula Eqn.\ref{eq:wellknown}. 
While higher-order spin contributions, such as 
  the interplay of the gravito-electric effect 
  with the GM effect (first appearing at $2.5$pN order)
  and spin-spin coupling (appearing at the $2$pN order),
  do influence the precession speed difference 
  between retrograde and prograde orbits, 
  they do not qualify as part of the GM clock effect. 
Therefore, we do not include the contributions of 
  these higher-order effects in this work.

%In general relativity, both the                                
%mass and centre-of-mass depend on the observer. 
%Similarly, the spin, 
%  being an angular momentum, 
%  is also affected by the choice of observer. 
  
The \ac{ssc} and, hence, the coefficient $k$ depend
  on the definition of the centre-of-mass.
For a spinning sphere of mass $m$ and spin magnitude ${\boldsymbol s}$, 
  the centre-of-mass depends on the observer \citep[see e.g.][]{Costa2012}. 
For an observer co-moving with the spinning sphere, 
  the centre-of-mass coincides with its geometrical centre.
While for an observer moving with ${\boldsymbol v}$ 
  with respect to the sphere, 
  the centre-of-mass is displaced by ${\boldsymbol v} \times {\boldsymbol s} /m$.
The set of centres-of-mass determined 
  by all possible observers 
  form a disk perpendicular to the spin axis with a radius of $s/m$, 
  which is called the M\"{o}ller radius.

In principle, any observer can be chosen to describe the motion of the spinning object, 
  although some choices are more natural than others. 
In the literature, there are a                                         % DS
few conventional choices which give rise to different values for $k$:
  the Frenkel-Mathisson-Pirani condition \citep{Frenkel1926,Mathisson1937}, which is also equivalent to the Tulczyjew-Dixon \citep{Tulczyjew1959,Dixon1964} condition at linear order in spin,
  corresponds to the case of $k=1$.
At                                                            
linear order in spin, 
  these two conditions both correspond to adopting an observer 
  co-moving with the spinning particle.
The Newton-Wigner-Pryce condition 
\citep{Pryce1948,Newton1949}
 is associated with $k=1/2$. 
Although the physical interpretation of this condition remains unclear,
  it has certain advantages. 
For instance, it is the only \ac{ssc} that leads to 
  vanishing Poisson brackets in both flat spacetime \citep{Pryce1948,Newton1949,Hanson1974}
  and curved spacetime \citep{Barausse2009}.   
The Corinaldesi-Papapetrou condition \citep{Corinaldesi1951},
  which is equivalent to adopting a set of static observers,          % DS
  corresponds to $k=0$.                                               % DS

The general-relativistic representation of these \acp{ssc} for a spinning particle with 4-momentum $p^{\mu}$
  and spin tensor $S^{\mu \nu}$ may be written as follows:
%\begin{align}
%{\setstretch{1.4}
%\begin{array}{lcr}
%S^{\mu \nu} u_{\nu} = 0 \,\; {\rm or} \,\; S^{\mu \nu} p_{\nu} = 0 &  \implies & k  = 1 \, , \\
%S^{\mu \nu } (p_{\nu} - m e^{\hat{0}}{}_{\nu}) = 0  & \implies & \hspace*{1mm} k = \frac{1}{2} \, , \\
%S^{0 \nu} = 0  & \implies & k = 0  \, .
%\end{array}}
%\end{align}
\begin{subequations} 
\begin{eqnarray}
S^{\mu \nu} u_{\nu} = 0 \,\; {\rm or} \,\; S^{\mu \nu} p_{\nu} = 0 & \hspace{0.5em} \implies \hspace{0.5em} & k  = 1 \, , \\
S^{\mu \nu } \left( p_{\nu} - m e^{\hat{0}}{}_{\nu} \right) = 0  & \hspace{0.5em} \implies \hspace{0.5em} & k = \frac{1}{2} \, , \\
S^{0 \nu} = 0  & \hspace{0.5em} \implies \hspace{0.5em} & k = 0  \, .
\end{eqnarray} 
\end{subequations}
 
Here $m \equiv \sqrt{- p^{\mu} p_{\mu}}$ is the dynamical mass 
  and $u^{\mu}$ is the 4-velocity of the centre-of-mass of the spinning particle.   
The tensor $e^{\hat{a}}{}_{\nu}$ represents the natural tetrad field satisfying 
\begin{align}
e^{\hat{a}}{}_{\mu} e^{\hat{b}}{}_{\nu} g^{\mu \nu} = \eta^{\hat{a} \hat{b}} \ ,
\end{align}
where $\eta^{\hat{a} \hat{b}} = {\rm diag} (-1, 1, 1, 1)$. 
The tensor field $e^{\hat{0} {\nu}}$ is time-like and past-oriented,       
  such that $e^{\hat{0}}{}_{0} > 0$.

%%%%%%%%%%%%%%%%%%%%%%%%%%%%%%%%%%%%%% 
%%%%%%%%%%%%%%%%%%%%%%%%%%%%%%%%%%%%%% 
% Sec 3
\section{Perturbation to Newtonian orbits} 
\label{sec:perturbation}

\subsection{Perturbation to the orbit}
\begin{figure}
%	  \vspace*{0.05cm}   \center 
%          \vspace*{6cm}          
    \centering
    \vspace*{-0cm}
    \includegraphics[width=1\columnwidth]{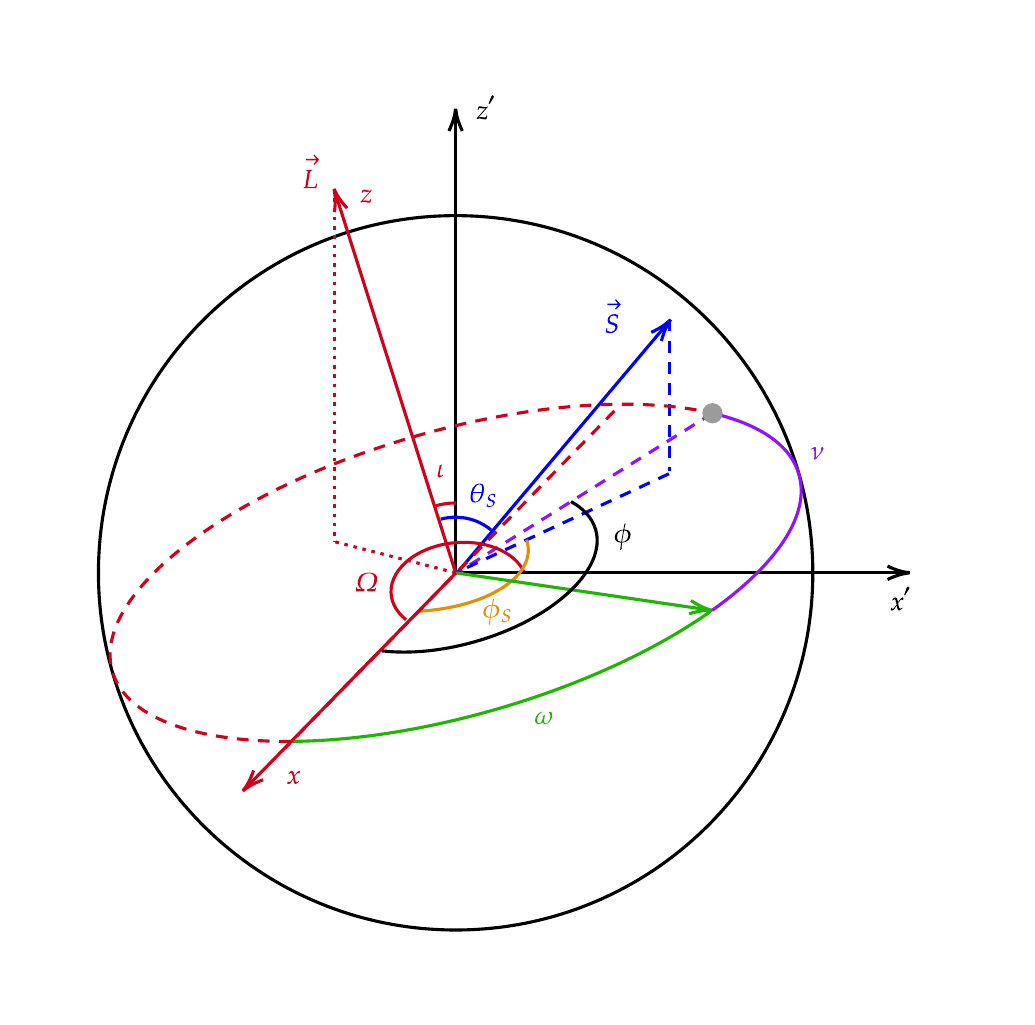}
    \vspace*{-0.5cm}
    \caption{An illustration of the geometrical configuration of the system. 
    The unperturbed orbit is located on the $z=0$ plane in $(x,y,z)$ coordinates,         
     with its angular momentum in the ${\hat {\boldsymbol z}}$ direction. 
    $\omega$ is the argument of periapsis.
    $\phi$ is  the azimuthal angle with respect to ${\hat {\boldsymbol x}}$
    and $\nu \equiv \phi - \omega$ is the true anomaly.
    The observer is located on the $x^{\prime}-y^{\prime}$ plane. 
    $\iota$ and $\Omega$ 
      are the inclination angle and longitude of ascending node, 
      respectively, with respect to $x^{\prime}-y^{\prime}$ plane. 
    The dashed lines represent lines on and projections onto the $x-y$ plane.
    The dotted lines represent projections onto the $x^{\prime}-y^{\prime}$ plane. 
    The polar angle $\theta_{S}$  
       and azimuthal angle $\phi_{S}$  
       of the total spin vector are defined with respect to the                   
    $(x,y,z)$ coordinate.
    The mass-weighted spin vector, which is similarly defined, is omitted from the figure.
    }
    \label{fig:geometry}
\end{figure}

In this section, we follow the method in \citet{Mashhoon2001} to 
  calculate the orbital perturbation due to the spin-orbit couplings 
  for a binary system. 
In the absence of spin, the binary's motion 
  follows a Newtonian orbit 
  confined to the plane perpendicular to the orbital angular momentum. 
We refer to the orbit of a non-spinning binary as 
  the \textit{unperturbed} orbit and the orbit of a slowly-spinning binary as the \textit{perturbed} orbit. 
Let us first consider the Cartesian coordinates $(x,y,z)$
  with ${\hat {\boldsymbol z}}$ parallel to the unperturbed orbital angular momentum. 
The total spin ${\boldsymbol S}$ has polar angle $\theta_{S}$ and azimuthal angle $\phi_{S}$ with respect to the $(x,y,z)$ coordinate. Similarly, the mass-weighted spin has polar angle $\theta_{\sigma}$ and azimuthal angle $\phi_{\sigma}$. 
The geometry of the binary system in shown in Fig.\ref{fig:geometry}. 

%%%%%%%%%%%%%%%%%%%%%%%%%%%%%%%%%%%%%%%%%%%%%%%%%% 
% Figure 1 
%%%%%%%%%%%%%%%%%%%%%%%%%%%%%%%%%%%%%%%%%%%%%%%%%%  

It is convenient to adopt cylindrical coordinates $(\rho, \phi, z)$ 
  for the orbit, 
  with the ${\hat {\boldsymbol z}}$-axis parallel to the (unperturbed) 
  orbital angular momentum. 
In the following, the subscript ``$0$'' denotes the unperturbed Newtonian 
  orbit. 
We have $z_0=0$ for the                                                  % DS
unperturbed orbit. 
The other two coordinates are given by:
\begin{subequations}
\begin{align}
\rho_0 &= \frac{a_0 (1-e_0^2)}{e_0 \cos \nu_0 +1}  \, , \label{eq:rho0} \\
\phi_0 &=  \omega_0 + 2 \arctan{\left[ \sqrt{\frac{1+e_0}{1-e_0}  } 
 \;\! \tan{\frac{E_0}{2}} \right] } \, ,  \label{eq:phi0}
\end{align}
\end{subequations}
where 
$a_0$ and $e_0$ are the (unperturbed) semi-major axis 
  and eccentricity, 
  $\nu_0 \equiv \phi_0-\omega_0$ is the true anomaly, 
  and $E_0$ is the eccentric anomaly, 
  which is related to the orbital period $P_0$ 
  via $E_0 - e_0 \sin{E_0} = 2 \pi/{P_0}$.  
For a Newtonian orbit,   
  $P_0 = 2 \pi \sqrt{{a_0}^3/M}$. 
The angular momentum of the unperturbed orbit per reduced mass is $L_0 \equiv \sqrt{M a_0 (1 - {e_0}^2)}$.

Cylindrical coordinates are related to Cartesian coordinates via
%  $x \equiv \rho \cos{\phi}$, $y \equiv \rho \sin{\phi}$ and $z \equiv z$
$(x,\, y,\, z) \equiv (\rho \cos{\phi},\, \rho \sin{\phi}, \, z)$.
We may therefore write the spins ${\boldsymbol \sigma}$ and ${\boldsymbol S}$ in cylindrical coordinates, thereby decomposing the acceleration due to the spin-orbit coupling force into the three directions:
\begin{align}
{\boldsymbol a}_{\rm SO} & =
C_{\rho} \frac{\dot{\phi }}{\rho^2} {\hat {\boldsymbol \rho}} 
+ C_{\phi} \frac{\dot{\rho}}{\rho^3} {\hat {\boldsymbol \phi}}  +
\left( C_{z,\phi} \frac{ \dot{\phi } }{\rho^2}+C_{z,\rho} \frac{\dot{\rho} }{\rho^3} \right) {\hat {\boldsymbol z}} \ ,
\end{align}
where
\begin{subequations}
\begin{align}
C_{\rho} &=  3 k \sigma  \cos{\theta _{\sigma }}+2 S \cos{\theta _S} \, , \\
C_{\phi} &= 3 (k-1) \sigma  \cos{\theta _{\sigma }}-2 S \cos{\theta _S} \, , \\
C_{z,\phi} &= 3 \sigma  \sin{\theta _{\sigma }} \cos (\phi -\phi _{\sigma })+4 S \sin{\theta _S} \cos (\phi -\phi _S) \, , \\
C_{z,\rho} &= 3 (k-1) \sigma  \sin{\theta _{\sigma }} \sin (\phi -\phi _{\sigma })-2 S \sin{\theta _S} \sin (\phi -\phi _S) \, , 
\end{align}
\end{subequations}
Therefore, the \ac{eom} of the binary under the perturbation is given by
\begin{subequations}
\begin{align}
\ddot{\rho}-\rho \dot{\phi}^{2}+\frac{ M}{\rho^{2}} &= C_{\rho} \frac{\dot{\phi }}{\rho^2} \label{eq:eom1} \, , \\
\rho \ddot{\phi}+2 \dot{\rho} \dot{\phi} &= C_{\phi} \frac{\dot{\rho}}{\rho^3} \label{eq:eom2} \, , \\
\ddot{z}+\frac{M}{\rho^{3}} z &= C_{z,\phi} \frac{ \dot{\phi } }{\rho^2}+ \epsilon \, C_{z,\rho} \frac{\dot{\rho} }{\rho^3} \label{eq:eom3} \, . 
\end{align}
\end{subequations}
We have introduced a new factor, $\epsilon \equiv |{\boldsymbol a}_{\rm SO}|/|{\boldsymbol a}_{\rm N}|$,
to quantify the magnitude of the spin-orbit coupling force 
  relative to the Newtonian gravitational force.
The expressions on the right hand sides of the Eqns.~\eqref{eq:eom1}--\eqref{eq:eom3} are of order $\sim \mathcal{O}(\epsilon)$.                            % DS
Since we are primarily interested in the \ac{gm} clock effect 
  under the assumption of slow rotation, 
  we retain only the terms up to first order 
  in the perturbation due to the spin-orbit coupling force. 
Consequently, all terms of order $\mathcal{O}(\epsilon^2)$ or higher are ignored.
Equation \eqref{eq:eom2} can also be rewritten as
\begin{align} 
 \frac{ {\rm d} (\rho^2 \dot{\phi})} {{\rm d}\;\! t}   
  =  \frac{C_{\phi} \dot{\rho}}{\rho^2} \, , 
\end{align} 
 whose solution is simply 
\begin{align} \label{eq:phiSolution}
  \rho^2 \dot{\phi} = C_1 -  \frac{C_{\phi}}{\rho} \, ,
\end{align}
where $C_1$ is some constant which should reduce to the angular momentum $L_0$ when both spins vanish. 
%                                                                       % DS
%                                                                       % DS
Therefore, we have $C_1 - L_0 \sim \mathcal{O}(\epsilon)$, 
  $\rho - \rho_0 \sim \mathcal{O}(\epsilon)$, and
  similarly for $\phi$ and $z$. 

%                                                                       % DS
  
Using the formula for $\dot{\phi}$ in Eqn.~\eqref{eq:phiSolution}, 
  and defining $u \equiv 1/\rho$, Eqn.~\eqref{eq:eom1} can be rewritten, with the help of Eqn.~\eqref{eq:rho0}, as a second order differential equation for $u$:
%%
%\begin{align} \label{eq:uSecondDifferential}
%\frac{{\rm d}^2 u}{{\rm d} \phi^2} + u &  =  \frac{M}{C_1^2} + 
%\frac{1}{a_0^2 C_1 \left(1-e_0^2 \right)^2} \bigg\{  \left(C_{\phi }-C_{\rho}\right) \left[ e_0 \cos \left(\phi_0 -\omega _0\right)+1\right]^2  \nonumber \\
%& \hspace*{0.5cm} 
%  +C_{\phi } \left[ 1-e_0^2 \cos \left( 2 \phi_0 -2 \omega _0\right)\right]  \bigg\} 
%  \ , 
%\end{align}
%%
%
\begin{equation}
\begin{aligned} \label{eq:uSecondDifferential}
\frac{{\rm d}^2 u}{{\rm d} \phi^2} + u = \frac{M}{C_1^2} &+ 
\frac{\left(1-e_{0}^2\right)^{-2}}{a_0^2 C_1} \bigg\{ C_{\phi } \left[ 1-e_0^2 \cos \left( 2\phi_0 - 2\omega_0\right)\right] \\
& + \left(C_{\phi }-C_{\rho}\right) \left[ e_0 \cos \left(\phi_0 -\omega _0\right)+1\right]^2
    \bigg\} 
  \ , 
\end{aligned} 
\end{equation}
where $C_1$ remains undetermined. 
The second term on the right hand side of Eqn.~\eqref{eq:uSecondDifferential}
  corresponds to a linear-order correction in $\epsilon$. 
Consequently, we may introduce a quantity similar to the semi-major axis,
  denoted by $a = a_0 ( 1+ \alpha)$ where $\alpha \sim \mathcal{O}(\epsilon)$,
  and a quantity similar to eccentricity, denoted by  
  $e = e_0 (1 + \eta)$ where $\eta \sim \mathcal{O}(\epsilon)$.
We can use these definitions interchangeably with $a_0$ and $e_0$ 
  as per convenience, 
  and this substitution would only contribute to the perturbation 
  at the second order in $\epsilon^2$.
We can also replace $C_1$ with $L_0$ using the same reasoning.
  
Combining this differential equation and form of unperturbed orbit, we can guess the solution to be of the form:
\begin{align} \label{eq:uPerturbed}
u = \frac{1+ e \cos \left[ \left(1+ C_3 \right) \phi  -\omega \right]+  C_2   \cos \left[ 2 \left(  1 +  C_3\right) \phi  -2 \omega \right]}{a \left(1-e^2\right)} \, ,
\end{align}
where $\omega \equiv \omega_0 +  \xi$.
It is important to note that these definitions do not 
  necessarily represent the semi-major axis, 
  eccentricity, and argument of periapsis of the perturbed orbit. 
However, these definitions correspond to those 
  of the unperturbed orbit 
  when both spins vanish.
In the solution, we have $C_1 - L_0 \sim \mathcal{O}(\epsilon)$. 
$C_2$, $C_3$, $\alpha$, $\eta$ and $\xi$ are all small corrections 
  of linear order in $\epsilon$.
Among these, $C_3$ represents the orbital precession due to the spin-orbit coupling force,
  while $C_2$ represents the non-sinusoidal variation of the orbit. 
There are five                                                        % DS
unknowns ($a, e, \xi, C_2, C_3$)                                      % DS
in this solution, and one                                             % DS
unknown $C_1$ in the differential equation. 
Next, we perform a Taylor expansion of the proposed solution
  with respect to $\epsilon$ and 
  compare it with the differential equation, 
  disregarding terms of order $\mathcal{O}(\epsilon^2)$.
The constant term in $u$, 
  and the 
  coefficients of $\phi  \sin ({\phi }-\omega_0)$ 
  and $\cos( {2 \phi} - 2 \omega_0)$ 
  inside the Taylor expansion of $u$  
  would be used to eliminate three  
  degrees of freedom, leaving $\alpha$, $\eta$  
  and $\xi$ as the free variables.
It follows that 
\begin{align}
C_2 & = \frac{e_0^2 M }{6 L_0^3 }  \left(C_{\rho}+C_{\phi }\right) \, ,
\end{align}
\vspace*{-3mm}
\begin{align}
C_3 & = \frac{M }{L_0^3 }  \left(C_{\rho}-C_{\phi }\right) \, ,
\end{align}
\vspace*{-3mm}
\begin{equation}
\begin{aligned}
C_1 = L_0 -\frac{  M }{4 L_0^2} 
 & \bigg\{ 2 a_0 L_0 \left[ e_0^2 (\alpha +2 \eta )-\alpha \right]    \\
&  \hspace*{1cm}  + (e_0^2+2 ) C_{\rho} - (e_0^2+4 ) C_{\phi } \bigg\} \, . 
\end{aligned}
\end{equation}
The remaining free variables can be determined by the value of 
  $u$,  $\dot{u}$ and 
  $\dot{\phi}$ at  a given time instance 
  (or at three different time instances).

It is worth noting that $C_1$ can also be expressed as a function of $a$ and $e$:
\begin{align}
C_1 & = \sqrt{M a (1- e^2) } + 
\frac{1}{4 a \left(1- e^2\right)} \left[ \left(e^2+4 \right) C_{\phi } - \left(e^2+2 \right) C_{\rho} \right] \, .
\end{align}
The motion in the $z$-direction can be solved by assuming $z=\rho H(\phi)$ \citep{Mashhoon1978,Mashhoon2001}, 
where $H(\phi)$, as well as its derivatives, 
  are of linear order in $\epsilon$.  
Ignoring second order perturbations, 
 and applying Eqns.~\eqref{eq:eom1}--\eqref{eq:eom3}, 
  we obtain a differential equation for $z$:
\begin{align}
\rho \dot{\phi}^2 \left(\frac{{\rm d}^2 H}{{\rm d \phi}^2 } + H \right) & =   C_{z,\phi} \frac{ \dot{\phi } }{\rho^2}+  C_{z,\rho} \frac{\dot{\rho} }{\rho^3} \ .  
\end{align} 
(Note that both sides of the equation are linear in $\epsilon$.)   
Replacing $\rho$, $\dot{\phi}$, and $\dot{\rho}$ with $1/u$, 
  $\left(C_1 u^2 - \epsilon  C_{\phi} u^3\right)$, and $-\dot{u}/u^2$, respectively, and upon ignoring terms above first order, the differential equation for $H(\phi)$ may be rewritten as:
\begin{align}
\frac{{\rm d}^2 H}{{\rm d \phi}^2 } + H & = \frac{ M}{L_0^3}   \bigg\{  e_0 C_{z,\rho} \sin (\phi -\omega _0)+C_{z,\phi} [ e_0 \cos (\phi -\omega _0)+1] \bigg\} \, .
\end{align}
The general solution of $H(\phi)$
  that is accurate to linear order in $\epsilon$ is:
%%
%\begin{align}
%H =
%%& \frac{\epsilon  e_0 M }{2 L_0^3} \bigg\{ 3 k \sigma \sin{\theta_{\sigma}} \cos \left(\phi_{\sigma} -\omega_0\right)+2 S \sin{\theta_S} \cos \left(\phi_S-\omega_0\right) \\
%%& - 2 S \sin{\theta_S} \cos \left[ 2 \phi -\left(\omega_0+\phi_S\right)\right] \\
%%& + (k-2) \sigma \sin{\theta_{\sigma}} \cos \left[ 2 \phi -\left(\omega_0+\phi_{\sigma} \right)\right]  \bigg\} \\
%%&+\frac{\sigma \sin{\theta_{\sigma}}}{L_0^2} \cos \left[  \left(1-\epsilon  \frac{ 3 M }{2 L_0}  \right) \phi  -\phi_{\sigma} \right] \\
%%&+ \frac{S \sin{\theta_S}}{L_0^2} \cos \left[\left(1-\epsilon  \frac{ 4 M }{2 L_0}  \right)\phi   -\phi_S\right] 
%%+ C_4 \cos{\phi} + C_5 \sin{\phi}
%%\ , \\
%& \frac{M}{4 L_0^3}
%\bigg\{ 
%\phi  \Big[6 \sigma  \sin \theta_{\sigma } \sin \left(\phi -\phi_{\sigma }\right)+8 S \sin \theta_S \sin \left(\phi -\phi_S\right) \Big] \\&  + C_4 \sin \phi  
%  +C_5 \cos \phi  
% +2 e_0 \Big[
%\sigma  \sin \theta_{\sigma }  \big(3 k \cos \left(\phi_{\sigma }-\omega_0\right)   \\ 
%&  + (k-2)  \cos \left(2 \phi-\phi_{\sigma }-\omega_0 \right)\big) 
%\\ 
%& +4 S \sin \theta_S \sin \left(\phi -\phi_S\right) \sin \left(\phi -\omega_0\right) \Big]
%\bigg\} \ ,
%\end{align}
%%
%
\begin{equation}
\begin{aligned}
H =
& \frac{M}{4 L_0^3}
\bigg\{ 
\phi  \Big[6 \sigma  \sin \theta_{\sigma } \sin \left(\phi -\phi_{\sigma }\right)+8 S \sin \theta_S \sin \left(\phi -\phi_S\right) \Big] \\&  + C_4 \sin \phi  
  +C_5 \cos \phi  
 +2 e_0 \Big[
\sigma  \sin \theta_{\sigma }  \big(3 k \cos \left(\phi_{\sigma }-\omega_0\right)   \\ 
&  + (k-2)  \cos \left(2 \phi-\phi_{\sigma }-\omega_0 \right)\big) 
\\ 
& +4 S \sin \theta_S \sin \left(\phi -\phi_S\right) \sin \left(\phi -\omega_0\right) \Big]
\bigg\} \, ,
\end{aligned}
\end{equation}
with $C_4$ and $C_5$ being arbitrary constants
  which are linear in $\epsilon$. 
These two degrees of freedom
  can be determined by the value of $z$ and 
  $\dot{z}$ at a given time instance (or two different instances) .
Further, it is notable that $H$ grows secularly as $\phi \to \infty$, 
  emphasising that the formula will be less accurate as the number 
  of orbit increases.

\subsection{Projection of orbit on the observer's coordinate system}

We adopt a new Cartesian coordinate system 
  $(x^{\prime},\,y^{\prime},\,z^{\prime})$ 
  for the observer (i.e., a terrestrial radio telescope in case of pulsar timing or a laser ranging site for artificial satellite) with the observer placed on the $x^{\prime}-y^{\prime}$ plane. 
We note that the definition of inclination angle here
  is different from the convention adopted in binary system observations,
  and an edge-on orbit corresponds to $\iota = 0 $ in our definition.   
To simplify calculations, we may ignore the relative motion between 
  the centre-of-mass of the binary and the observer, 
  such that both coordinate systems $(x,\,y,\,z)$ and $(x^{\prime},\,y^{\prime},\,z^{\prime})$ remain static.  
These coordinate systems are related via the rotation:
\begin{equation}
\left[\begin{array}{c}
x^{\prime}  \\ y^{\prime} \\ z^{\prime}
\end{array}\right]
= \left[
\begin{array}{ccc}
\cos \Omega  & -\cos \iota  \sin \Omega  & \sin \iota  \sin \Omega  \\
 \sin \Omega  & \cos \iota  \cos \Omega  & -\cos \Omega  \sin \iota  \\
 0 & \sin \iota  & \cos \iota  \\
\end{array}
\right]
\left[\begin{array}{c}
x \\ y \\ z
\end{array}\right] \ , 
\end{equation}
where $\Omega$ is the longitude of the ascending node and $\iota$ is the inclination angle with respect to the reference plane $x^{\prime}-y^{\prime}$.
We also introduce the spherical coordinate system $(R,\,\varphi,\,\vartheta)$, which is defined with respect to $(x^{\prime},\,y^{\prime},\,z^{\prime})$.

\subsection{General solution}
Because $a_0$, $e_0$ and $\omega_0$ are all parameters                      % DS
that are not directly measurable, the results are instead expressed in terms of $a$, $e$ and $\omega$. 
This choice of variables eliminates 
  the three degrees of freedom represented by $\alpha$, $\eta$, and $\xi$.
Due to precession, the orbit is no long closed 
  (expect for when the orbit is strictly circular, even under perturbations). 
To account for this, we introduce 
  an additional term $\phi_1$, of linear order in $\epsilon$,
  which characterises the nutation and precession of the orbital plane.
Assuming a full orbit starts at $\phi = \phi_0$ and ends at
  $\phi = \phi_0 + 2\pi + \epsilon \phi_1$, 
  the time it takes to complete a full orbit is:
\begin{equation}
\begin{aligned} \label{eq:genericclock}
\mathcal{T} & = \int_{\phi_0} ^ {\phi_0 + 2 \pi + \epsilon \phi_1 } \frac{{\rm d} \phi} { C_1 u^2 - \epsilon C_{\phi} u^3 } \\
& =2 \pi \sqrt{\frac{a^3}{M}} +
\frac{\pi}{2 (1-e^2 )^{5/2} M} \, \times \\
& \hspace*{17mm} \big[ (-3 e^4+3 e^2-2 ) C_{\rho} + (3 e^4-9 e^2+4 ) C_{\phi } \big] \\
& \hspace*{14mm} +\frac{1}{M \left[ 1 + e \cos \left(\omega -\phi _0\right)\right]^2} \, \times \\
& \hspace*{17mm} \big[  2 \pi   \left(C_{\rho}-C_{\phi }\right) + a^{3/2} (1-e^2 )^{3/2} \sqrt{M} \, \phi_{1} \big] \, .
%& =2 \pi \sqrt{\frac{{a_0}^3}{M}}  \left(1 + \frac{3}{2} \alpha \right) + 
%\frac{1}{2 (1-{e_0}^2 )^{5/2} M} 
%\bigg\{  \pi  \big[ (-3 {e_0}^4+3 {e_0}^2-2 ) C_{\rho} \\
%& +(3 {e_0}^4-9 {e_0}^2+4 ) C_{\phi } \big] + 
%\frac{2 (1-{e_0}^2)^{5/2} }{\left[ 1 + e \cos \left(\omega -\phi _0\right)\right]^2} \bigg\{ \big[ 2 \pi   \left(C_{\rho}-C_{\phi }\right) \big] \\
%& + {a_0}^{3/2} (1-{e_0}^2 )^{3/2} \sqrt{M} \phi _1 
%\bigg\}
\end{aligned}
\end{equation}
The angle $\phi_1$ depends on the definition of a full orbit. 
To linear order in $\epsilon$, the period difference between the perturbed and unperturbed orbit is given by:
\begin{align}\label{eq:genericclock}
\Delta \mathcal{T}&  = \mathcal{T} -  2 \pi \sqrt{\frac{{a_0}^3}{M}}  =  \Delta \mathcal{T}_{\rm size} + \Delta \mathcal{T}_{\rm fixed} 
+ \Delta \mathcal{T}_{\rm orb}  \ , 
\end{align}
where
\begin{subequations}
\begin{align}\label{eq:genericclock}
\Delta \mathcal{T}_{\rm size} & = 2 \pi \sqrt{\frac{{a}^3}{M}} - 2 \pi \sqrt{\frac{{a_0}^3}{M}} = 3 \pi   \alpha \sqrt{\frac{{a_0}^3}{M}}  \ , \\
\Delta \mathcal{T}_{\rm fixed} & = \frac{\pi 
 \big[ (-3 e^4+3 e^2-2 ) C_{\rho} +(3 e^4-9 e^2+4 ) C_{\phi } \big]}{2 (1-e^2 )^{5/2} M} 
   \ ,  \\
\Delta \mathcal{T}_{\rm orb} &  = 
 \frac{\big[   a^{3/2} (1-e^2 )^{3/2} \sqrt{M} \phi_1 
 + 2 \pi   \left(C_{\rho}-C_{\phi }\right)  
\big]}{M \left[ 1 + e \cos \left(\omega -\phi _0\right)\right]^2}  \ . 
%\Delta \mathcal{T}_{\rm fixed} & = \frac{\pi}{2 (1-e^2 )^{5/2} M} 
%  \big[ (-3 e^4+3 e^2-2 ) C_{\rho} +(3 e^4-9 e^2+4 ) C_{\phi } \big]  \ ,  \\
%\Delta \mathcal{T}_{\rm orb} & = \frac{1 }{M \left[ 1 + e \cos \left(\omega -\phi _0\right)\right]^2}  \nonumber \\ 
% & \times   \big[   a^{3/2} (1-e^2 )^{3/2} \sqrt{M} \phi_1  + 2 \pi   \left(C_{\rho}-C_{\phi }\right)  \big]  \ .   \\ 
\end{align}
\end{subequations}
To linear order in $\epsilon$, replacing $e$ with $e_0$ does not affect the three terms, hence they are independent of the free parameter $\eta$. 
$\Delta \mathcal{T}_{\rm fixed} $ is independent of the choice of $\alpha$, $\phi_1$ or $\phi_0$. 
$\Delta \mathcal{T}_{\rm size}$ depends on the free parameter $\alpha$, 
  and therefore represents the period difference due to the difference 
  in size of the perturbed and unperturbed orbits. 
As we will discuss later, this difference is related to the non-unique definition of ``nearly-identical'' orbits. 
$\Delta \mathcal{T}_{\rm orb}$ depends on the definition of                  % DS
a full revolution and on the initial position $\phi_0$. 
Because $\alpha$ and $\phi_1$ are free parameters 
  (as long as they are linear in $\epsilon$), $\Delta \mathcal{T}_{\rm fixed} $ and $\Delta \mathcal{T}_{\rm orb}$ are somewhat arbitrary and can themselves vanish, or be fine-tuned to make $\Delta \mathcal{T}$ vanish under suitable parameter choices. 

This is part of the reason why the \ac{gm} clock effect formula differs across the literature.                                            % DS
These differences will be discussed in further detail                                         % DS 
in the following section. 
To clarify, the \ac{gm} clock difference usually
  refers to the period difference between a prograde orbit 
  and a retrograde orbit. 
In our case, we define $\Delta \mathcal{T}$ as the difference in period 
  between a spinning and non-spinning binary system,
  which is about half of the 
  \ac{gm} clock effect commonly referred to in the literature.

\subsection{A full revolution: $\Delta \mathcal{T}_{\rm orb}$}
\label{sec:fullrevolution}
\begin{figure}
    \centering
    \vspace*{-0cm}
    \includegraphics[width=0.9\columnwidth]{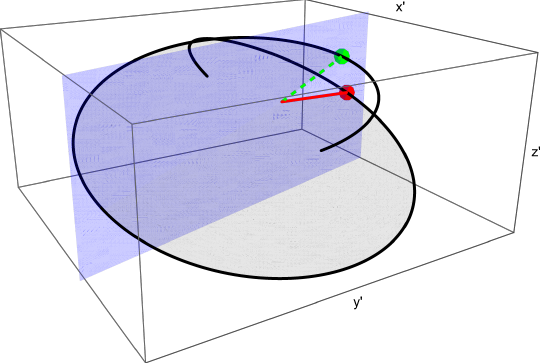}
    \includegraphics[width=0.9\columnwidth]{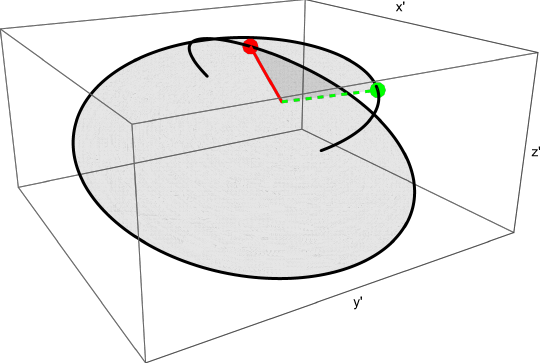}
    \includegraphics[width=0.9\columnwidth]{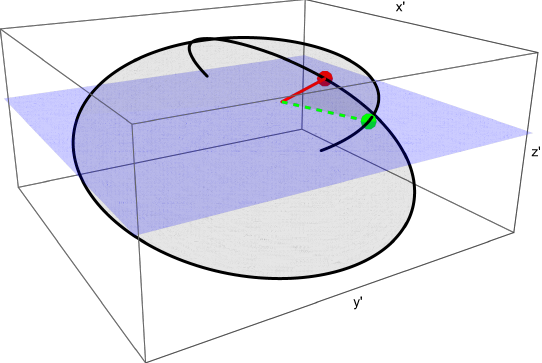}
    \vspace*{-0.0cm}
    \caption{These figures illustrate different ways of defining a full orbital period. 
    The top panel shows the sidereal period defined with respect to a distant   % DS
    observer 
    on the vertical plane. 
    The middle panel shows the anomalistic                                      % DS
    period, which is independent of observer's position. 
    The bottom panel displays the nodal period, which is the duration between two 
    successive upward crossings of the $z'=0$ plane.
    In each panel, the black solid line represents the perturbed orbit, 
      with the perturbing force exaggerated for clarity.
    The green and red points indicate the start and end points of the specific orbital period, respectively, and are connected to the origin with green dashed and red solid lines.
    The sector that the orbit sweeps through during one orbital period is shaded with grey colour. 
    }
    \label{fig:fullOrbit}
\end{figure}

Consider an observer at  
  $(R, \varphi, \vartheta) = (\infty, \varphi_0, \pi/2)$ and 
  a binary approximately edge-on (such that $\iota \ll \pi/2$).
A full orbit can be defined as the orbit with $\varphi$  moving from $\varphi_0$ (which corresponds to $\phi =\phi_0$) to $\varphi_0+2 \pi$ (which corresponds to $\phi =\phi_0 + 2\pi + \epsilon \phi_1 $), with $\varphi$ defined as:
\begin{align}
\tan{\varphi} \equiv \frac{y^{\prime}}{x^{\prime}} = \frac{\sin \Omega \cos \phi+\cos \Omega \cos \iota \sin \phi-\epsilon H(\phi) \cos \Omega \sin \iota }{\cos \Omega \cos \phi-\sin \Omega \cos \iota \sin \phi+\epsilon H(\phi) \sin \Omega \sin \iota } \ .
\end{align}
Such an azimuthal closure would correspond to two sequential pulsar 
  superior conjunctions (or approximately, two maximum Shapiro delays)                              % DS
  for this observer.
The duration of such an orbit 
  is known as the sidereal period, 
  as shown in the top panel of Fig.~\ref{fig:fullOrbit}.
With this definition, the angle $\phi_1$ can be calculated by solving  $\tan{\varphi} |_{\phi = \phi_0} = \tan{\varphi} |_{\phi = \phi_0+ 2\pi + \epsilon \phi_1} $, yielding:
%%
%\begin{align}\label{eq:azimuthalClosure}
%\phi_1 = & - \frac{\pi   M}{{L_0}^3} \  
%  \bigg[\; 3 \sigma \sin{\theta_{\sigma}} \sin \left(\phi_{\sigma}  
%  -\phi_0\right)    \nonumber  \\
%&  \hspace*{1.2cm}
%+4 S \sin{\theta_S} \sin \left(\phi_S -\phi _0\right) \bigg] \ 
%   \tan \iota \cos \phi _0   \ .
%\end{align}
%%
%
\begin{equation}
\begin{aligned}\label{eq:azimuthalClosure}
\phi_1 = & - \frac{\pi   M}{{L_0}^3} \tan \iota \cos \phi_{0} \, \times \\ 
 & \hspace*{3.5mm} \bigg[\; 3 \sigma \sin{\theta_{\sigma}} \sin \left(\phi_{\sigma}  
  -\phi_0\right) + 4 S \sin{\theta_S} \sin \left(\phi_S -\phi _0\right) \bigg] \, .
\end{aligned}
\end{equation}
This formula is independent of the \ac{ssc} condition as it roots from the 
  ${\boldsymbol v} \times (3 {\boldsymbol \sigma} + 4 {\boldsymbol S})$
  part of the spin-orbit coupling force,
  from which the parameter $k$ is absent.
When the  inclination angle is zero, the $x-y$ plane overlaps with the
  $x^{\prime}-y^{\prime}$ plane and we have $\phi_1 = 0$, as expected. 

%%%%%%%%%%%%%%%%%%%%%%%%%%%%%%%%%%%%%%%%%%%%%%%%%% 
% Figure 2
%%%%%%%%%%%%%%%%%%%%%%%%%%%%%%%%%%%%%%%%%%%%%%%%%%

There are other possible definitions of a full orbit. 
Considering an observer on the $x^{\prime}-y^{\prime}$ plane, the azimuthal closure is 
  well defined for nearly edge-on orbits, but it loses its meaning for face-on orbits.
A more sensible definition of a full revolution 
  for eccentric face-on orbits can                                                % DS
  be defined by the time interval between two successive periapsis passages, 
  which is also known as the anomalistic period,
  as shown in the middle panel of Fig.~\ref{fig:fullOrbit}.
This definition corresponds to the time difference 
  between two sequential maxima of the Einstein time delay. 
Under this assumption, the full orbit is 
  defined as ${\rm d} u / {\rm d}\phi |_{\phi = \phi_0} = {\rm d} u / {\rm d}\phi |_{\phi = \phi_0 + 2\pi + \epsilon \phi_1} = 0$. 
Both the initial position $\phi_0$, and $\phi_1$, are independent of the observer's location. The expression for $\phi_1$ is given by:
\begin{align}\label{eq:anomalisticphi1}
\phi_1 =     \frac{ 2 \pi M }{{L_0}^3 }  \left(C_{\phi }-C_{\rho}\right)  \, .
\end{align}   
That this definition of a full revolution yields $\Delta \mathcal{T}_{\rm orb} = 0$ is expected.
%
%\begin{align}
%\Delta \mathcal{T}_{\rm orb} = 0 \ .
%\end{align}
%
Although this expression is not divergent as $e_0 \to 0$, 
  it is important to emphasise that it is not applicable to circular orbits.
% DS 
In the limit of a                                                            % DS 
circular orbit, the physical meaning of the                                     % DS
anomalistic period becomes invalid due to the absence of periapsides. 
The bottom panel of Fig.~\ref{fig:fullOrbit} demonstrates the nodal period (or draconic period).
It is defined as the time interval between 
  two successive upward crossings of the $x^{\prime}-y^{\prime}$ plane. 
These crossing points are known as the ascending nodes, 
  which correspond to the vanishing of
\begin{align}
z^{\prime} = \epsilon \rho  H(\phi ) \cos \iota + \rho  \sin \iota \sin \phi  \, .
\end{align}
For simplicity, we set $H(0) = 0$ such that the first ascending node
  corresponds to $\phi = 0$. 
The second passage can be found by solving $z^{\prime}|_{\phi = 2 \pi + \epsilon \phi_1} = 0$.
To linear order in $\epsilon$, one finds:
\begin{align}
\phi_1 & = 
\frac{\pi  M }{{L_0}^3} \cot \iota  \left(3 \sigma  \sin \phi_{\sigma} \sin \theta _{\sigma } +4 S \sin \phi_S \sin  \theta_S \right) \, . 
\end{align}
As expected, the nodal period diverges when $\iota \to 0$, since the ascending node is not well defined in this case. 
When $\iota = \pi/2$ we have $\phi_1 = 0$. 

The above definitions are not unique. 
For example, the synodic period, which is defined
  with reference to two or more observers (e.g., the Earth and the Sun)
  would be a more feasible definition for the measurement of the \ac{gm} clock effect 
  with artificial satellites.   
%\citet{Iorio2016} calculated the gravito-electric effects on the different periods while 
Most importantly, all of these different definitions of orbital period 
  are identical for a binary under 
  Newtonian gravity in the absence of any perturbing force, but differ for perturbed orbits.

\subsection{Nearly-identical orbits: $\Delta \mathcal{T}_{\rm size}$}
\label{sec:nearlyidentical}
% There are no identical orbit
The                                                                            % DS
\ac{gm} clock effect is usually represented by the difference in the orbital period of a prograde (i.e. ${\boldsymbol S}\cdot {\boldsymbol L} >0$) orbit and that of an identical retrograde (i.e., ${\boldsymbol S}\cdot {\boldsymbol L} <0$) motion. 
However, from a theoretical perspective, there are no identical orbits when the          % DS
spin-orbit force is included. 
This is simple to illustrate: the radial motion, i.e., Eqn.~\eqref{eq:uPerturbed}, has additional terms with approximately two                                                                               % DS
times the orbital frequency (i.e., the term with coefficient $C_2$) and a precession term, $C_3$.
While it is possible to fix both $C_2$ and $C_3$ to vanish simultaneously, the motion in the $z$-direction cannot also be fixed to vanish, and vice versa. 
Even for circular orbits which seem very much similar, the angular velocity for orbits with identical radii will be different. 
This difference in angular velocity is in fact the source of the \ac{gm} clock effect.

% we need to define nearly identical orbits 
Therefore, before proceeding to calculate the \ac{gm} clock effect, the orbits that we choose to compare must be carefully specified. 
We will refer to these orbits as ``nearly-identical'' or indistinguishable orbits.  
% alpha
The non-unique choice of what constitutes 
  two ``nearly-identical'' orbits can be translated into the choice of the free variable $\alpha$, which is somewhat arbitrary and can be fine-tuned to make the time difference vanish. 
For example, if we define two ``nearly-identical'' circular orbits as two circular orbits with the same 
 $\dot{\phi}$ but slightly different radii (at the $\sim \mathcal{O}(\epsilon)$ level), the clock effect for a full revolution will vanish for these two nearly-identical orbits.

The definition of ``nearly-identical'' is best clarified in terms of observable parameters. 
For pulsar observations, the main observables are the time delay due to pulse-arrival-time and Doppler effects. 
% A general definition
In general, it would be natural to define nearly-identical orbits as those that result in indistinguishable observational signatures.
For example, if two different orbits can be fitted by the same template with exactly the same parameters, these two orbits will be considered nearly-identical.
% The problem with details
However, such a definition can be quite subtle, and is dependent on the observation carried out, the noise level, and the template used to fit the data.

% different time delays
In the simplified model where the centre-of-mass of the binary is at rest with respect to the observer, there are several sources of time delays, in particular:
(i) the Roemer delay due to the displacement of the pulsar relative to the centre-of-mass, (ii) the Shapiro delay due to the passage of light ray around the companion, and (iii) the Einstein delay due to time dilation in the gravitational field. 
As for the                                                                            % DS
Doppler effect, its importance depends on the magnitude of the pulsar's velocity projected onto the line-of-sight. 
Consequently, different measurements are sensitive to different orbital parameters.

For example, the Roemer delay is sensitive to the (projected) size of the orbit.  
Two orbits of the same projected size but with slightly different velocities will be indistinguishable from the point of view                                                              % DS
of the Roemer delay.
The Einstein delay depends on the variation of the gravitational potential (i.e., $m_{2}/r$) from the companion star.
Roughly speaking, this effect depends on the eccentricity and semi-major axis of the orbit. 
Hence, two orbits with the same eccentricity and same semi-major axes will be considered as nearly-identical orbits if only the             % DS
Einstein time delay is used to determine the binary parameters. 
The                                                                                   % DS
Doppler effect depends on the projected velocity and is therefore not as sensitive to the shape of the orbit.

\begin{figure}
    \centering
    \vspace*{-0.2cm}
    \includegraphics[width=1\columnwidth]{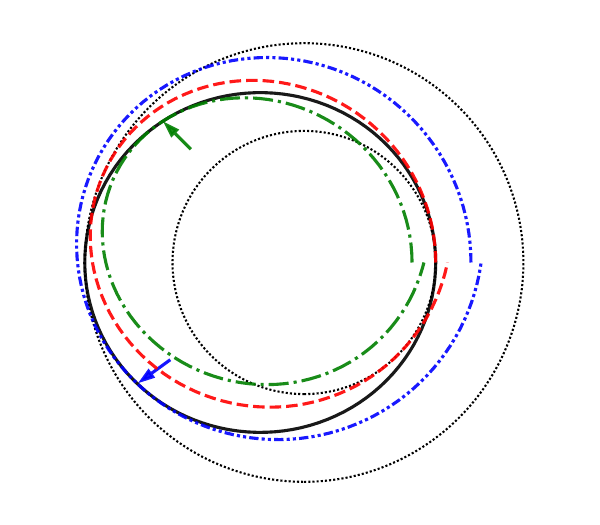}
    \vspace*{-1.0cm}
    \caption{This is a representation of                                         % DS
    different definitions of nearly-identical perturbed orbits. 
    The black solid ellipse represents the unperturbed orbit. 
    The two dotted circles represent circular orbits with radii                  % DS
    $r_{\min}$ and $r_{\max}$. 
    The semi-ellipses represent perturbed orbits which are nearly-identical to the ellipse under different definitions. 
    The red dashed semi-ellipse represents a perturbed orbit with the same $r_{\min}$ and $r_{\max}$ as 
    the                                                                          % DS
    unperturbed orbit.     
    The green dashed-dotted semi-ellipse and the blue dashed-dotted-dotted semi-ellipse
      represent                                                                  % DS
      an orbit that connects to the unperturbed orbit
      smoothly at $\phi = 0.75 \pi$ and $\phi = 1.2 \pi$, respectively. 
    Each of the connecting points is highlighted with an arrow.    
    The perturbation is exaggerated (i.e. $\epsilon \sim 0.15$) 
    to assist visualisation. }
    \label{fig:nearlyidentical}
\end{figure}

% Now combine all the info
In practice, all of these different time delays are combined to estimate the best fit orbital parameters. 
Whether two orbits are distinguishable or not depends on the binary parameters, the orientation of the orbit, and the noise level, which introduce random factors into the estimated parameters. 
A detailed and observationally-consistent definition of 
 ``nearly-identical'' orbits is complex and is beyond the scope of this work. 
In the following, we will use a few simplified assumptions to demonstrate how the definition of nearly-identical orbits affects the value of the \ac{gm} clock effect. 
We will consider a few typical examples which are presented in Fig.~\ref{fig:nearlyidentical}.

%%%%%%%%%%%%%%%%%%%%%%%%%%%%%%%%%%%%%%%%%%%%%%%%%% 
% Figure 2
%%%%%%%%%%%%%%%%%%%%%%%%%%%%%%%%%%%%%%%%%%%%%%%%%%

A system dominated by the                                                          % DS
Einstein delay would consider orbits
  with the same  $r_{\min}$ and $r_{\max}$ as nearly-identical. 
Because $r = \sqrt{\rho^2 + z^2}$, the oscillation in the $z-$direction only contributes to changes in the radial distance at second order in $\epsilon$.
Therefore, we have $r_{\min} \approx \rho_{\min} = u_{\max}$ and similarly for $r_{\max}$.   
This requires 
\begin{align}
\alpha & = \frac{C_2 (1+{e_0}^2 )}{1-{e_0}^2} \ , \\
\eta & = C_2 \ . 
\end{align}
Using this relation in the case of $\Delta \mathcal{T}_{\rm orb}=0$
  (i.e. the anomalistic period),  
  the clock effect between the period of the perturbed and that of the unperturbed orbit is given by
\begin{equation}
\begin{aligned}
\Delta \mathcal{T}_{\rm size} + \Delta \mathcal{T}_{\rm fixed}& = -\frac{\pi  \left(C_{\rho}-2 C_{\phi }\right)}{M \sqrt{1-{e_0}^2}}  \\ 
& = \frac{3 \pi 
  \left[ (k-2) \sigma  \cos \theta _{\sigma } -2 S \cos  \theta _S \right]}{M\sqrt{1-{e_0}^2} }  
 \, . 
% & = \frac{3 \pi  }{M\sqrt{1-{e_0}^2} }  \left[ (k-2) \sigma  \cos \theta _{\sigma } -2 S \cos  \theta _S \right] \ .  \nonumber   \\ 
\end{aligned}
\end{equation}
Since the coefficients of $\sigma$ and $S$ are both negative (for $k = 0, 1/2, 1$), 
  this implies that when $\Delta \mathcal{{T}}_{\rm orb} =0$,
  it takes less time for a binary to complete a full revolution on a prograde orbit
  than on an unperturbed orbit or a perturbed retrograde orbit. 
This counter-intuitive conclusion is specific 
  to the anomalistic period.
As spin-orbit coupling forces cause the orbit to precess, 
  the periaspsis vectors rotate in the opposite direction 
  of the angular velocity for prograde orbits. 
Therefore, an anomalistic period is complete for $\Delta \phi < 2 \pi$ on a 
  prograde orbit, so that the time it takes to finish a full revolution is in fact shorter
  than the corresponding unperturbed or retrograde orbit. 

Another example is given by a system that connects the                        % DS
perturbed orbit and unperturbed orbit smoothly. 
If we assume the unperturbed orbit turns                                      % DS
into a perturbed orbit at $\phi = \phi_0$, it is equivalent to assuming that the spin-orbit coupling force 
is turned on                                                                  % DS
at $\phi = \phi_0$. 
%At $\phi = \phi_0$, we require $u (\phi=\phi_0)$, ${\rm d}u/{\rm d} t (\phi=\phi_0)$, ${\rm d} \phi/{\rm d} t (\phi=\phi_0)$ of the perturbed orbit coincide with those of unperturbed orbit. Similarly, we have  $z(\phi=\phi_0)= 0  = {\rm d}z/{\rm d} t (\phi=\phi_0)$. 
%
%         % DS:  I rewrote the sentence that's blocked above to remove the excessive use of \phi = \phi_0.
%
We require that $u$,  $\dot{u}$, and $\dot{\phi}$  
of the perturbed orbit at $\phi = \phi_0$ coincide with their counterparts of the unperturbed orbit. 
Similarly, we have   $z = \dot{z} = 0$   at $\phi = \phi_0$. 
This requires
\begin{equation}
\begin{aligned}
\alpha = & \phantom{+} \frac{a_0 M^2 }{6 {L_0}^5} 
\bigg\{
2 e_0 \left( C_{\rho}+C_{\phi }\right) \big[ 2 {e_0}^2 \cos ^3 (\phi _0-\omega _0 ) \\
& + 3 e_0 \cos (2 \phi _0-2 \omega _0 ) + 6 \cos  (\phi _0-\omega _0 )\big] \\
& + 3 \left({e_0}^4+{e_0}^2+2\right) C_{\rho}-3 \left({e_0}^2-5\right) {e_0}^2 C_{\phi}
\bigg\} \ . 
\end{aligned}
\end{equation}
We omit the formula of the                                                  % DS
\ac{gm} clock effect under this condition as it is redundant.

\section{Comparison with previous studies}                                  % DS
\label{sec:literature} 

In this section, we compare the \ac{gm} clock effect we defined 
  with the results from \citet{Mashhoon2001}, \citet{Bini2004a,Bini2004b}, \citet{Mashhoon2006} and \citet{Hackmann2014}. 
The systems they studied are all \ac{emri} systems, with special focus on the Earth-satellite system.
For an \ac{emri} system, we consider $m_1$ to be the mass of the test particle and $m_2$ to be the massive object. This system could refer to a pulsar orbiting a massive black hole or a satellite orbiting the Earth. 
Therefore, we have $m_1/m_2 \ll 1$ and $M \equiv m_1 + m_2 \approx m_2$. 
To leading order in the mass ratio,
  we have ${\boldsymbol S} \approx {\boldsymbol S}_2 \sim m_2^2$
  and ${\boldsymbol \sigma} = m_2 {\boldsymbol S}_1/m_1 + m_1 {\boldsymbol S}_2/m_2 \sim m_1 m_2$.
We note that for the
  Earth-satellite system ${\boldsymbol \sigma}= 0$ due to the negligible spin of the satellite.
Furthermore,                                                                % DS
we note that 
  in \citep{Bini2004a,Bini2004b,Mashhoon2006}, they have used 
  ${\boldsymbol \sigma} = m_2 {\boldsymbol S}_1/m_1$ instead,
  such that the clock effect formula is accurate to                         % DS
  leading order in mass ratio
  for ${\boldsymbol S}_1$ and ${\boldsymbol S}_2$, respectively.

In addition, the                                                            % DS
\ac{eom} for the test particles are                                          % DS
usually described as geodesics (with or without perturbations due to the  spin of the small mass  ${\boldsymbol S}_1$) in the Kerr spacetime, 
  and the inertial frame is usually defined 
with reference to ${\boldsymbol S}_2$, instead of the orbital angular momentum. 
The observer's frame $(x^{\prime},\, y^{\prime},\, z^{\prime})$ is defined such that ${\boldsymbol S}$ overlaps with the                                                             % DS
  $z^{\prime}$-axis. 
Therefore, the inclination angle of the orbital plane $\iota$ is equivalent to $\theta_{S}$. 
This difference in definition encodes                                         % DS
implicit restrictions of the parameter space and 
 naturally leads to discrepancies in the results, as illustrated in the following subsections.

\subsection{Mashhoon's clock effect}

\citet{Mashhoon2001} investigated 
the clock effect of a test particle orbiting the Earth. 
The mass and spin of the Earth dominates the spin-orbit coupling forces. 
Therefore, the mass-weighted 
spin vector ${\boldsymbol \sigma}$ and                                     % DS
the spin ${\boldsymbol S}_1$ can be ignored. 
For the observer on the Earth, its $z'$ axis matches the spin axis, i.e., $\theta_S = \iota$ and $\phi_{S} = \pi/2$. 
The two                                                                   % DS
degrees of freedom in the  $H(\phi)$ are removed by requiring $H(\phi_0) = H^{\prime} (\phi_0) = 0$. 
With these conditions, the coefficients of the forces are given by
\begin{subequations}
\begin{align}
  C_{\rho}  &= 2 S \cos \iota \, , \\
  C_{\phi } &= -2 S \cos \iota \, , \\
 C_{z,\rho} &= 2 S \cos \phi  \sin \iota \, , \\
 C_{z,\phi} &= 4 S \sin \phi  \sin \iota  \, . 
\end{align}
\end{subequations}
These conditions yield                                                    % DS
$C_2 =0$, where the orbital semi-major axis and eccentricity of the perturbed orbit coincide with $a$ and $e$. 
The authors defined a full orbit using the
  sidereal period, i.e., Eqn.~\eqref{eq:azimuthalClosure}.
This is equivalent to fixing the angle $\phi_1$ to be
\begin{align}\label{eq:phi1Mashhoon}
\phi_1 & = -\frac{4 \pi  M }{{L_0}^3} S \sin{\iota} \tan {\iota } \cos^2 {\phi_0} \ .
\end{align}
\citet{Mashhoon2001} commented that $\iota$ should not be near $\pi/2$ to ensure $\phi_1$ remains small, given the divergence of $\tan \iota$ around $\iota = \pi/2$. 
We seek to provide a more precise range of the required inclination angle.
For a satellite in low Earth orbit, to ensure that $\phi_1 \ll 1$, we must have
%
% 4*pi*mass of Earth^2*0.33*(2*pi/1day)*(6400 km)^2*Sqrt(G)/(c^2 * Sqrt(6400 km*mass of Earth)^3) 
\begin{equation}
\begin{aligned}
\frac{\pi}{2} - \iota & \gg \frac{4 \pi M_{\oplus} I_{\oplus}\omega_{\oplus}}{{\left(a  M_{\oplus}\right)}^{3/2}} \\ 
& = 1.7 \times 10^{-10} \,{\rm rad}\,
\left( \frac{I_{\oplus}}{8 \times 10^{37}\,{\rm kg}\,{\rm m}^2} \right)
% \left( \frac{1\,{\rm day}}{P_{\oplus}} \right)
\left( \frac{6400\,{\rm km}}{a} \right)^{{3}/{2}}
\, ,
\end{aligned}
\end{equation}
where we have used $I_{\oplus} \simeq 0.33 M_{\oplus} {R_{\oplus}}^2$ 
  to calculate the Earth's moment of inertia, with 
  the coefficient $0.33$ taken from \citet{Williams1994}.
This condition implies that Eqn.~\eqref{eq:phi1Mashhoon}
  remains valid for the majority of existing satellites,
  suggesting that the clock effect could 
  potentially be revealed with existing data. 
The time it takes to complete an azimuthal closure is therefore
\begin{align}\label{eq:Mashhoon1}
\mathcal{T} & =  2 \pi \sqrt{\frac{a^3}{M}} + \frac{2 \pi S \cos \iota }{M} 
\left\{ \frac{4 - 2 \cos^2 \!\phi_0 \tan^2 \iota}{\big[ 1 + e \cos(\phi_0 - \omega )\big]^2 } - \frac{3}{\sqrt{1- e^2}} \right\} \, .
\end{align}
This result agrees precisely with Eqn.~(30) of \citet{Mashhoon2001}.

\subsection{Bini's results: circular equatorial orbits with spinning secondary}

\citet{Bini2004a,Bini2004b} studied the \ac{gm} clock effects in \acp{emri} with both perturbed and unperturbed orbits, which are strictly circular and restricted to the equatorial plane.
The unperturbed orbit, the perturbed prograde orbit, and the perturbed retrograde orbit, are all assumed to have the same radius.   
This corresponds to cases with $e = e_{0} = 0$ and $\alpha=0$, such that $a = a_{0}$ represents the constant distance between the binary pair.
Both spins are required to be perpendicular to the orbital plane (i.e., $\sin{\theta_{\sigma}} = \sin{\theta_S} = 0$) such that the motion in the $z$-direction vanishes. 
For a                                                                     % DS
strictly circular orbit, we have $\phi_1 = 0 $ regardless of the orientation of the observer's plane.
The time it takes for the binary to complete one orbit becomes
\begin{align}
%\mathcal{T} & =  2 \pi \sqrt{\frac{a^3}{M}} +  \frac{\pi  }{M}  \left( \pm 2 S \pm 3 k \sigma  \right)\, ,
\mathcal{T} & =  2 \pi \sqrt{\frac{a^3}{M}} +  \frac{2\pi  }{M}  \left( \pm \frac{3}{2} k \sigma \pm S  \right)\, ,
\end{align}
where the plus (minus) sign depends on the relative orientation of the spin axis and corresponds to the spin axis being aligned (anti-aligned) with the orbital angular momentum.
%For an \ac{emri} system, if we adopt ${\boldsymbol \sigma} \approx m_2 {\boldsymbol S}_1/m_1$ and ${\boldsymbol S} \approx {\boldsymbol S}_2$,
For an \ac{emri} system, assuming ${\boldsymbol \sigma} \simeq m_2 {\boldsymbol S}_1/m_1$ and ${\boldsymbol S} \simeq {\boldsymbol S}_2$,
the orbital time scale becomes
\begin{align}\label{eq:Bini}
%\mathcal{T} & =  2 \pi \sqrt{\frac{a^3}{M}} +  \frac{2 \pi  }{M}  \left( \pm S_2 \pm \frac{3}{2} k \frac{M}{m_1} S_1 \right)\ , 
\mathcal{T} & =  2 \pi \sqrt{\frac{a^3}{M}} +  \frac{2 \pi  }{M}  \left( \pm \frac{3}{2} k \frac{M}{m_1} S_1 \pm S_2 \right)\ , 
\end{align}
This result matches Eqns.~(3.2)--(3.3) of \citet{Bini2004b} for the three \acp{ssc} considered in their paper. It also matches the result of \citet{Faruque2004} when $k=1$.

\subsection{Mashhoon and Singh's results}

\citet{Mashhoon2006} studied the \ac{gm} clock effect for an \ac{emri} system on a semi-circular orbit close to the equatorial plane. 
The circular geodesics are considered as unperturbed orbits and the force due to the interaction of curvature with the spin of the test particle (${\boldsymbol S}_1$) is considered as the perturbing force. 
The spin of the massive object is therefore either aligned or anti-aligned with the orbital angular momentum, but there is no restriction on the orientation of the smaller spin.

The nearly-identical perturbed orbit is defined as the orbit that the test particle will follow if the force due to the smaller spin is switched on at $t=0$.
More specifically, two orbits with the same ${\boldsymbol S}$ but with either vanishing ${\boldsymbol S}_1$ or non-vanishing  ${\boldsymbol S}_1$, that share the same $u$, $\dot{u}$, $\phi$, $\dot{\phi}$, and $z=0=\dot{z}$ at $t=0$, are considered nearly-identical. 
The azimuthal closure is defined as $\phi_0 \to \phi_0 + 2 \pi$, such that the motion in the $z$-direction (and hence the projection of the spin onto the orbital plane) can be ignored and we have $\phi_1=0$. 
Under this condition, we have $e_0 = 0$ and
\begin{align}
\alpha & =  e_0 \eta = \frac{3 k \sigma }{\sqrt{{a_0}^3 M}}    \, . 
\end{align}
The unperturbed orbit is strictly circular but the perturbed orbit admits a small eccentricity. 
The time it takes for the smaller particle to complete one full orbit is
\begin{align}\label{eq:MashhoonSingh}
%\mathcal{T} & =  2 \pi \sqrt{\frac{{a_0}^3}{M}}  +  \frac{2 \pi  }{M}  \left( \pm  S_2  +  6 k \frac{M}{m_1} S_1 \cos{\theta_{S_1}} \right)\, .
\mathcal{T} & =  2 \pi \sqrt{\frac{{a_0}^3}{M}}  +  \frac{2 \pi  }{M}  \left( 6 k \frac{M}{m_1} S_1 \cos{\theta_{S_1}} \pm  S_2 \right)\, . 
\end{align}
The sign in front of $S_2$ depends on the relative orientation of the orbit. 
As the perturbed orbit remains close to the equatorial plane, the sign of $S_2$ is positive (negative) when the spin of the black hole is aligned (anti-aligned) with the orbital angular momentum. 
The angle $\theta_{S_1}$ is not restricted and can be written as ${\hat {\boldsymbol S}_1} \cdot {\hat {\boldsymbol J}}$.
\citet{Mashhoon2006} used the                                                % DS
Tulczyjew-Dixon condition, which suggests $k=1$.
Under this condition, our result becomes equivalent to Eqn.~(57) in their paper.
In comparison to the work of \citet{Bini2004a,Bini2004b} and \citet{Faruque2006}, 
  the coefficient of $S_1$ in \citet{Mashhoon2006} and Eqn.~\eqref{eq:MashhoonSingh} is much larger. 
We confirm that the difference in this coefficient is merely an artefact of the different literature definitions of ``nearly-identical'' orbits.
 
\subsection{Hackmann and L\"{a}mmerzahl's results}                          % DS
\label{sec:hackmann}
\citet{Hackmann2014} compared inclined orbits of a                          % DS
non-spinning test particle in the Kerr and Schwarzschild spacetimes.
The Kerr geodesics were considered to be the perturbed orbit while the Schwarzschild geodesics were the unperturbed orbit. 
The spin axis ${\boldsymbol S}={\boldsymbol S}_2$ 
  was aligned (or anti-aligned) with the $z$-axis, while the orbital angular momentum was inclined at an angle $\iota$ with respect to the $z$-axis. 
The orbit precesses and nutates at different frequencies around the spin axis, as a consequence of the spin-orbit coupling force. 
They calculated the clock effect by averaging the azimuthal velocity over radial oscillation and altitudinal                                          % DS
oscillation respectively.
As these two oscillations have different frequencies in the presence of perturbative forces, their result is equivalent to averaging the \ac{gm} clock effect over infinite time.
Further, by using the inverse azimuthal velocity to derive the clock effect, they have effectively considered the azimuthal closure condition and their formula represents the infinitely averaged sidereal period. 
Consequently, their result is independent of the argument of periapsis of the orbit. 
The formula, i.e., Eqn.~(30) of \citet{Hackmann2014}, in our notation reads:
\begin{align}\label{eq:Hackmann}
\Delta \mathcal{T}  & = \frac{2 \pi S}{M}  
\frac{  \left(3 e^2+2 e+3\right) \cos \iota -2 e-2 }{\left(1-e^2\right)^{3/2}} \, .
\end{align}
In general, spin-orbit coupling forces give rise to the                       % DS
\ac{gm} clock terms that contain $\theta_{S}$ (which is equivalent to $\iota$ in this case).
However, their formula has a component that is independent of $\iota$. 
We argue that this term is a direct consequence of the precession of the orbital angular momentum and that, although not explicitly stated, their averaging method has assumed the orbital angular momentum of the non-spinning test particle to be much smaller than ${\boldsymbol S}$.
When $S \gg \mu L$, the orbital angular momentum approximately precesses around the spin axis \citep{Kidder1995} with
\begin{align}\label{eq:orbitalLPrecession}
\dot{{\boldsymbol L}} & = \frac{2}{r^3}\;\! {\boldsymbol S}_2 \times {\boldsymbol L} \, .  
\end{align}
The prograde motion acquires a positive angular velocity such that $\omega_S = {2S_2}/{r^3}$, regardless of the inclination angle.
In fact, this angular velocity is the precession speed of the longitude of the ascending node (i.e., nodal precession) of the satellite \citep{Lense1918}.
The correction to the orbital period due to this angular velocity is 
\begin{align}
\Delta \mathcal{T} & = \frac{2 \pi}{\omega_{\rm N} + \omega_S} - \frac{2 \pi}{\omega_{\rm N}}
= -\frac{4 \pi  S}{M} + \mathcal{O}(S^2) \, , 
\end{align}
for a circular orbit, which is equivalent to the part of Eqn.~\eqref{eq:Hackmann} that is independent of the inclination angle when $e=0$. 
Here $\omega_{\rm N} = \sqrt{M/r^3}$ is the Newtonian angular velocity. 
This relation implies that the inclination-independent component of the clock effect originates from the nodal precession of the orbit. 
During the node's prograde precession with respect to the spin of the massive object, it contributes to reducing the orbital period of the prograde motion and increasing the orbital period of the retrograde motion.
As the azimuthal velocity has been averaged over infinite time, the clock effect was calculated with respect to a distant and fixed observer located on the $x^{\prime}-y^{\prime}$ plane.
This physical picture suggests that the spinning axis is effectively fixed with respect to the distant observer while the orbital angular momentum precesses    % DS
about this axis. 
As both spin and orbital angular momenta precess around the total angular momentum, ${\boldsymbol J} = {\boldsymbol S}_2 + \mu {\boldsymbol L}$, this picture is only valid when ${\boldsymbol J} \approx {\boldsymbol S}_2$, which implies $|{\boldsymbol S}_2| \gg \mu |{\boldsymbol L}|$. 
This requirement suggests that the formula will be valid in the test particle limit (e.g., an Earth-satellite system) but may be violated in compact pulsar-black-hole systems, where the angular momentum of the pulsar 
is non-negligible compared with the black hole's spin.

In comparison, our perturbation method is valid regardless of the mass ratio and whichever angular momentum dominates. 
Since we have ignored precession of the orbital angular momentum and spin angular momenta, our formula is valid when the observational time scale is much shorter than the precession time scale. 
For this reason, it is unnecessary to compare our formula with the result of \citet{Hackmann2014}. 
Nevertheless, these two methods lead to consistent results when the spin axes are approximately aligned (or anti-aligned) with orbital angular momentum, as precession may be ignored in such cases.
When $\iota=0$, Eqn.~\eqref{eq:Mashhoon1} reduces to:
\begin{equation}
\begin{aligned}\label{eq:Hackmann2}
\Delta \mathcal{T}  & = \frac{2 \pi S   }{M}  \left<
 - \frac{3}{\sqrt{1- e^2}} + \frac{4}{\big[ 1 + e \cos(\phi_0 - \omega )\big]^2 }  
\right>_{\omega}  \\
& =  \frac{2 \pi S   }{M}\frac{ 1 + 3 e^2 }{\left(1-e^2\right)^{3/2}} \, ,
\end{aligned}
\end{equation}
where $\left< \cdots \right>_{\omega}$ represents averaging over $\omega$.
This is the same as Eqn.~\eqref{eq:Hackmann} for $\iota=0$. 

%\citep{Mikóczi2017} 
%\citep{Tartaglia2001} calculated the circle orbits of photon in 
%  opposite directions. 

\subsection{Further comments on the clock effect}
It has been shown that the differences in various results in the literature are due to the choice of \ac{ssc} (i.e., the value of $k$), the definition of ``nearly-identical'' orbits (i.e., the value of $\alpha$), and the definition of a full revolution (i.e., the value of $\phi_1$). 
For example, the constant coefficients in front of the secondary spin could differ by a factor of four                                                            % DS
in Eqn.~\eqref{eq:Bini} and Eqn.~\eqref{eq:MashhoonSingh}, even for the same \ac{ssc}, as a consequence of different definitions of ``nearly-identical'' orbits. 
In fact, the coefficients can be made arbitrarily large or small with suitable choices of $\alpha$ and $\phi_1$. 
Some coefficients appear in the literature more often than others, simply because their underlying definitions seem more natural for the system being considered, as compared with others.

We seek to                                                            % DS
further illustrate this effect by showing that even the well-known formula Eqn.~\eqref{eq:wellknown} admits variants.
Consider an Earth-satellite system with a                                    % DS
satellite moving on an unperturbed circular orbit on the equatorial plane of the Earth. 
We now define a                                                             % DS
nearly-identical perturbed orbit as one that connects to the circular unperturbed orbit smoothly at periapsis. 
For an unperturbed orbit with radius $a_0$, we then have
\begin{align}
    \alpha & =  \frac{C_{\rho}}{ {a_0}^{3/2}  M^{1/2} }  \, .
\end{align}
The \ac{gm} clock effect is now given by
\begin{align}
    \mathcal{T}_{+} - \mathcal{T}_{-} & = \frac{16 \pi  S}{M} \, .
\end{align}
To the best of our knowledge, only the constant coefficient $4 \pi$ appears in the literature. 
This is because it seems more natural to consider two orbits with the same size ($r_{\min}$ and $r_{\max}$) as being nearly identical.  
For pulsar timing, this is indeed the case as the Roemer delay is, in general, the dominant time delay (with order $\sim r/c$), while the other effects depend on $v/c$ (the Doppler effect), $(v/c)^2$ (the Einstein delay), or $G m_2 / c^3$ (the Shapiro delay). 
However, the Roemer delay is only sensitive to the projected size of the orbit, hence for binaries with nearly face-on orientations the Roemer delay will vanish and decoding other time delays will be necessary to acquire the full information of the orbital parameters in pulsar timing observations. 

%%%%%%%%%%%%%%%%%%%%%%%%%%%%%%%%%%%%%% 
%%%%%%%%%%%%%%%%%%%%%%%%%%%%%%%%%%%%%%
\section{Conclusions}
\label{sec:conclusion}
% generic
In this study, we have derived the \ac{gm} clock effects for a binary system with arbitrary mass ratio, eccentricity, inclination angle, and spin orientation.
%under the generic assumption that 
%  the binary orbital angular momentum is much larger than both spin for arbitrary relative orientation. 
The difference in orbital period between a spinning binary and a non-spinning binary is given by Eqn.~\eqref{eq:genericclock}.
% degree of freedom
The generic clock effect admits three degrees of freedom: the choice of \ac{ssc} (i.e., value of $k$) which comes from the dependence of spin-orbit coupling forces on the \ac{ssc}, 
  the definition of ``nearly-identical'' orbits (i.e., the value of $\alpha$), and the definition of a full revolution (i.e., the value of $\phi_{1}$).
We demonstrate how these various definitions of ``nearly-identical'' orbits 
  affect the clock effect 
  in Sec.~\ref{sec:nearlyidentical} and show how the definition of a full revolution affects   
  the clock effect in Sec.~\ref{sec:fullrevolution}. 

% nearly identical
When the spin-orbit force is present, 
  the orbit can no longer to be identical to orbits where the          
  spin-orbit coupling force is absent. 
This can also be explained by the fact that the spin is non-degenerate and cannot be mimicked 
  by a Newtonian non-spinning binary. 
Therefore, the \ac{gm} clock effects depend on the orbits being compared.
We refer to those orbits being compared as ``nearly-identical'' orbits 
  and argue that this definition should be observationally                               % DS 
  motivated. 
%It is therefore sensible to define ``nearly-identical'' orbits 
%  as the ones that are observationally indistinguishable. 
In Sec.~\ref{sec:nearlyidentical}, 
   we presented some examples 
   of possible definitions of ``nearly-identical'' orbits 
   that are relevant for pulsar timing observations and demonstrated
   that these would lead to different formulae for the \ac{gm} clock effects. 
  
%As for pulsar timing, the orbital parameters are deduced by measuring different components of time delays, which are sensitive to different orbital elements.
%For example, the Roemer delay is sensitive to the projected size of the orbit; the Doppler effect is sensitive to the velocity projected along the line-of-sight; 
%  the Einstein delay is more sensitive to the eccentricity. 
%It is also possible to define ``nearly-identical'' orbits as the perturbed and unperturbed orbits that connects smoothly, as if the spin-orbit force is turned on at a specific 
%  time. While this definition seems natural in terms of simulation, 
%  it does not necessarily make much sense for observation purpose. 
  
% full revolution
In the context of general relativity, 
  the definition of a full revolution is not unique, 
  unlike in Newtonian mechanics. 
This is due to the precession of the argument of periapsis, 
   which gives rise to at least two different definitions: 
the sidereal period and the anomalistic period. 
For nearly edge-on orbits, 
  the sidereal period corresponds to the time 
  interval between two total eclipses (or two successive maxima of the Shapiro delay). 
For nearly face-on orbits, 
  the anomalistic period 
  is a more sensible definition, 
  as it represents the time interval between successive maxima of the Einstein delay. 
It is also possible to define the orbital period 
  as the nodal period, 
  but the physical meaning of this definition is ambiguous. 
For binaries with an intermediate inclination angle, 
  finding a sensible definition 
  is complicated by the potential degeneracy between the orbital parameters.
%These different definitions coincide for Newtonian orbits and are different subject to perturbations. 
%We leave these to future study.  
%  as the wobbling of the orbital plane (i.e. nutation of orbital angular momentum)
%  cast additional 
%  provides an additional contribution to the azimuthal closure, 
%  when the orbit is inclined at a non-zero angle. 

% literature

We compare the existing expressions in the literature 
  (in Sec.~\ref{sec:literature})  
  and demonstrate that 
  the generic clock effect formula that we derive 
  can recover most of these results. 
We have identified the source of discrepancies in the literature 
  and explained these differences 
  using the three degrees of freedom mentioned earlier. 
Compared to previous works, 
  our generic clock effect formula 
  can be applied to a broader parameter space, 
  including, but not limited to, 
  extreme-mass-ratio binaries. 
This study thus provides a useful framework  
  for the investigation of GM clocks effects 
  and orbital dynamics 
  in a broad range of systems, 
  from artificial satellites around the Earth 
  to general astrophysical systems 
  containing two spinning compact objects in 
  close orbits around each other.

%We also comment on the difference between our work and starting from solving geodesics or non-geodesics in the Kerr spacetime. 

%%%%%%%%%%%%%%%%%%%%%%%%%%%%%%%%%%%%%% 
%%%%%%%%%%%%%%%%%%%%%%%%%%%%%%%%%%%%%%
\section*{Acknowledgements}

%We thank **** for discussions and for critically reading of the manuscript.
We are grateful to Prof Bahram Mashhoon for valuable in-dpeth discussions 
  on gravito-electromagnetism analogy and the associated gauge issues.
KJL is supported by a PhD Scholarship from the Vinson and Cissy Chu Foundation, and by a UCL MAPS Dean's Prize. 
ZY is supported by a UKRI Stephen Hawking Fellowship.
JT is supported by a UK STFC Research Studentship.
KW and JT acknowledge support from the UCL Cosmoparticle Initiative. 
This work is supported in part by a UK STFC Consolidated Grant awarded to UCL-MSSL. 
This research has made use of NASA's Astrophysics Data System (ADS),
Mathematica, and the matplotlib package.

%%%%%%%%%%%%%%%%%%%%%%%%%%%%%%%%%%%%%% 
%%%%%%%%%%%%%%%%%%%%%%%%%%%%%%%%%%%%%%
\section*{Data availability}

No new data were generated or analysed in support of this research.

%%%%%%%%%%%%%%%%%%%%%%%%%%%%%%%%%%%%%%%%%%%%%%%%%%  
%%%%%%%%%%%%%%%%%%%% REFERENCES %%%%%%%%%%%%%%%%%%
\bibliographystyle{mnras}
\bibliography{reference.bib}  
%%%%%%%%%%%%%%%%%%%%%%%%%%%%%%%%%%%%%%%%%%%%%%%%%%

%%%%%%%%%%%%%%%%% APPENDICES %%%%%%%%%%%%%%%%%%%%%

\appendix
\section{Equation of motion and gauge symmetry}
 The GM clock effect is a consequence of
  the analogy between electromagnetic and 
  gravitational field.
%  an analogy extensively explored by (add references here). 
%In this context, 
%The initial derivations of the GM clock effect 
It was initially derived 
  \citep[e.g.][]{Vladimirov1984,Mashhoon2001,Iorio2002} 
  from a Lorentz-force-like formula, with the 
  gravito-electric field arising from the point mass, 
  and the GM (gravito-magnetic) field taking the form of a magnetic dipole
  \citep[see e.g.][]{Braginskii1977,Thorne1988,Harris1991,Clark2000,Mashhoon2001b,Mashhoon2003,Iorio2023}.
%Notably, 
In this work we adopt the spin-orbit coupling representation,  
  similar to that in \citet{Barker1979}
  through Eqn.~\ref{eq:spinorbitforce},   
  an approach different from the approaches 
  that explicitly use the Lorentz-force-like formula. 
The difference in the two formula
  stems from the geometrical nature of gravity.
While gauge transformation does not interfere with the 
  coordinates in the electromagnetic theory, 
  gauge transformation explicitly changes the coordinates
  in general relativity,
  and the Lorentz-force-like formula appears a 
  specific gauge choice, commonly referred to as the (gravito-electromagnetic) GEM gauge. 
Consequently, there is some arbitrariness 
  in the definition of the GM field, 
  depending on the construction of the analogy and the adopted gauge-fixing conditions \citep[see e.g.][]{Clark2000,Mashhoon2001b}.
Owning to this subtlety, it is unsurprising that the spin-orbit coupling force 
  presented in this work, which is derived under the standard pN gauge and ADM gauge, 
  differs from that of the GEM gauge. 
These disparities can be reconciled through appropriate coordinate transformations. 
However, we are not aware of existing literature 
  that addresses this difference, 
  possibly due to the limited overlap between the two fields.
Since Newtonian gravity does not exhibit this subtlety, 
  the differences between these sets of coordinates become apparent, 
  at least at the 1pN order.

The intricacies of this coordinate issue become particularly relevant  
  when connecting a locally defined field 
  (e.g., the GM field) 
  with a globally defined concept (e.g., orbital size and eccentricity), 
  which are commonly employed in pulsar timing, satellite ranging, and similar observations.
To counter this ambiguity, gauge-invariant quantities like orbital frequency 
  and Detweiler's redshift have been widely adopted to avoid coordinate-dependent artefacts. 
It's important to note that classical GM clock effect, 
  as demonstrated in Eqn.~\ref{eq:wellknown}, 
  does not exhibit such coordinate-dependent artefacts, 
  owing to its topological characteristics
  and symmetrical nature. 
Despite variations arising from different schemes, 
  the same clock difference is consistently obtained without the need to clarify the difference 
  in coordinates. 

Aligning two artificial satellites in the exact opposite 
  yet the same orbit is a challenging task. 
Realistic measurements need to account for formulae that 
  incorporate orbital size and eccentricity information, 
  which, inevitably, are written in a coordinate-dependent manner. 
The appearance of eccentricity may raise concerns, especially given 
  the various different definitions of eccentricity \citep[see e.g.,][]{Loutrel2019}, 
  and the unconventional definition we have adopted. 
As the clock effect due to eccentricity is already a perturbation 
  at the 1.5pN level,  
  any alternative definition of eccentricity would 
  alter the clock effect at (at least) the 2.5pN order.  
Therefore, we can safely ignore when considering leading-order effects.
  
In contrast, the semi-major axis presents a more serious challenge, as discussed earlier. 
We attempt to address this by introducing the somewhat arbitrary 
  parameter $\alpha$ in Eqn.~\ref{eq:genericclock}.
Exploring the intricate differences in coordinates defined in various gauges 
  is beyond the scope of this work. 
Instead, our focus is on directing the audience's attention to 
  the observationally oriented definitions, as argued in Sec.~\ref{sec:fullrevolution}
  and Sec.~\ref{sec:nearlyidentical}.

%%%%%%%%%%%%%%%%%%%%%%%%%%%%%%%%%%%%%%%%%%%%%%%%%%

% Don't change these lines
\bsp	% typesetting comment
\label{lastpage}
\end{document}